\documentclass[authoryear]{interact}
\usepackage{caption}
\usepackage[font=small,labelfont=bf]{caption}
\usepackage{graphicx}

\usepackage{newunicodechar}
\usepackage{hyperref}
\usepackage{array}
\usepackage{multirow}
\usepackage{longtable}
\usepackage{caption}
\usepackage[font=small,labelfont=bf]{caption}
\newunicodechar{，}{,}

\usepackage{epstopdf}
\usepackage[caption=false]{subfig}

\usepackage{url}
\usepackage{float}

\usepackage[sort]{natbib}
\bibpunct[, ]{(}{)}{;}{a}{,}{,}
\usepackage[T1]{fontenc}
\usepackage{lmodern}
\usepackage[utf8]{inputenc}
\usepackage{booktabs}
\usepackage{multirow}
\usepackage{tabularx}
\usepackage{adjustbox}

\theoremstyle{plain}

\theoremstyle{definition}

\theoremstyle{remark}

\begin{document}


\title{CoDesignAI: An AI-Enabled Multi-Agent, Multi-User System for Collaborative Urban Design at the Conceptual Stage}
\author{
\name{Zhaoxi Zhang $^\ast$  \textsuperscript{a}\thanks{$^\ast$ Corresponding Author: Zhaoxi Zhang, Department of Urban and Regional Planning, University of Florida. Email: zhang.zhaoxi@ufl.edu}, Ruolin Wu \textsuperscript{b}, Feiyang Ren \textsuperscript{b,c}, Sridevi Turaga\textsuperscript{b} and  Tamir Mendel \textsuperscript{d}}
\affil{
\textsuperscript{a}Department of Urban and Regional Planning, College of Design, Construction and Planning, University of Florida;
\textsuperscript{b}Center for Urban Science + Progress, Tandon School of Engineering, New York University, Brooklyn, 11201, USA;
\textsuperscript{c}School of Geography, University of Leeds, UK;
\textsuperscript{d}School of Information Systems, The Academic College of Tel Aviv-Yaffo, Tel Aviv, Israel;
}
}

\maketitle

\begin{abstract}

Public participation has become increasingly important in collaborative urban design; yet, existing processes often face challenges in achieving efficient and scalable citizen engagement. To address this gap, this study explores how large language models (LLMs) can support cooperation among community members in participatory design. We introduce CoDesignAI, a collaborative urban design tool that combines multiple users, representing residents or stakeholders, with multiple AI agents, representing domain experts who provide facilitation and professional knowledge during the conceptual stage of urban design. This paper presents the system architecture and main components of the tool, illustrating how users interact with AI agents within a collaborative and iterative design workflow. Specifically, the system integrates generative AI with spatial mapping services to support street-level visualization of design proposals. AI agents assist users by summarizing discussion content, extracting shared design intentions, and generating prompts for presenting design interventions. The system also enables users to revise and refine their ideas over multiple rounds while documenting the design process. By combining conversational AI, multi-user interaction, and image-based design grounded in real-world urban contexts, this study argues that AI-enabled design systems can help shift urban design from an expert-centered practice to a more open and participatory process. The paper contributes a new web-based platform for AI-assisted collaborative design and offers an early exploration of how AI agents may expand the capacity for public participation in urban design.

\end{abstract}

\begin{keywords}
Digital tool; Generative AI; Urban Design; Large Language Model; Citizen Science
\end{keywords}

\section{Introduction}

In urban planning and design, participatory approaches allow citizens to contribute local knowledge, lived experiences, and community concerns that are often overlooked by traditional planning methods, thereby supporting the co-creation of planning and design \citep{moore2016listening}. Public participation is a foundational concept in the theory of the “ladder of citizen participation”, developed by \citet{Arnstein1969}. This theory describes a progression from nonparticipation (a lower level of participation) to citizen power (a higher level of participation)\footnote{The ladder of citizen participation has eight rungs, progressing from low to high levels of public engagement: (1) Manipulation, (2) Therapy, (3) Informing, (4) Consultation, (5) Placation, (6) Partnership, (7) Delegated Power, and (8) Citizen Control.}. However, many empirical studies suggest that participation in practice often remains limited to informing and consultation, where the public’s influence over final decisions is still constrained \citep{kotus2013position,silverman2020we}. Although some studies have explored more collaborative forms of participation, real-world practice often reveals a persistent gap: government officials tend to prioritize procedural engagement, while the public often seeks deeper and more meaningful involvement in decision-making \citep{erdtman2025between}.


Since 2023, the rapid development of large language models (LLMs) and generative AI has created new opportunities for public service and community engagement \citep{Senadheera01022024,Henman01102020}. Recently, LLMs have been increasingly used to support more interactive forms of communication between institutions and the public, including public opinion collection \citep{li2026exploring} and participatory processes \citep{zhang2025leveraging,zhang2025ai}, due to their strong conversational abilities. In urban research and practice, this potential has attracted growing attention, and an increasing number of studies have begun to explore the usage of LLMs for participatory decision-making, design, and planning \citep{zhang2025leveraging,zhang2025ai,du2024artificial,zhang2023deliberating}. However, to the best of our knowledge, limited research has examined how human–AI collaboration can be integrated into participatory processes to support real-time, collaborative interactions among multiple users and multiple AI agents.

By leveraging LLMs and generative AI, this study explores the design of an AI-enabled multi-agent, multi-user system to support participatory design processes. As a proof of concept, this paper presents CoDesignAI, a web-based prototype platform that brings together diverse community members (i.e., multiple users) and AI expert agents (i.e., multiple agents) to explore new possibilities for urban design workflows and strengthen meaningful public participation. Rather than proposing a finalized solution, this study focuses on design exploration through prototype development, demonstrating how such a system may contribute to a more structured, interactive, and collaborative approach to urban design decision-making.

\section{Related Work}

\subsection{Participatory Urban Design}

Over the decades, urban designers and researchers have increasingly emphasized the participatory urban design process by promoting public participation throughout the entire project, from pre-design analysis and planning to final implementation and beyond \citep{Owens2024,Geekiyanage2021,Innes2004}. Traditionally, urban design has largely followed a top-down, expert-led approach in which public participation is often introduced only at the final stage of the process. In contrast, this new paradigm recognizes public participation as a critical force for encouraging co-creation across different stages of urban design \citep{foroughi2023public,richardson2013reinventing,butzlaff2023consenting}. To reveal the ``why” of co-creation in urban design, \cite{oetken2025unravelling} identified three key motivations for co-creation: generating societal impact through equity, creating strategic impact through innovation and adaptability, and fostering collaborative synergy through stakeholder engagement and collective knowledge exchange. Overall, participatory urban design is not only an urgent need in current practice but also an important direction for the future of urban design.

In practice, several methods and tools have been used to support participatory urban design with communities and citizens, including surveys \citep{shin2022designing}, public hearings \citep{conrad2011hearing}, interviews \citep{kearney2004walking}, focus groups \citep{kotchetkova2008articulating}, and community planning events \citep{wates2013community} to enhance public participation. Additionally, research has shown that public displays are useful for supporting public participation in urban settings \citep{du2017public}. In recent years, web-based geo-questionnaires, particularly public participation GIS (PPGIS), have been widely used to gather location-based public responses in urban settings \citep{amegbor2025spatial}. So far, different participatory methods in urban design each have distinct strengths and limitations. For example, interviews and focus groups can encourage deeper engagement by facilitating interaction with stakeholders and other participants \citep{Lambert2008}. However, these methods typically require considerable time and commitment from both participants and organizers, and those who participate are often highly motivated individuals or selected representatives \citep{baxter2023increasing}. The high time cost can also make it difficult to retain participants for follow-up studies. Surveys, especially online surveys, can reach a much larger number of people more efficiently and at a lower cost. However, they often face challenges related to biases concerning representativeness and overall data quality \citep{singh2021critical,szolnoki2013online}. In practice, different methods are often integrated \citep{kourkouridis2024urban,amegavi2025advancing}. Nevertheless, researchers continue to advocate for new participatory approaches that strike a better balance between meaningful engagement, efficiency, and inclusiveness.

In recent years, the rapid advancement of new digital tools has created opportunities to support and enhance human participation in urban design \citep{hanzl2007information}, as these tools represent an emerging paradigm that is shifting conventional workflows toward more interactive and collaborative processes \citep{turken2021participatory,toukola2022digital}. Current research identifies two main categories of digital tools: one involves device based tools, such as using virtual reality (VR) and augmented reality (AR) to compare the differences between two alternative design approaches in the design process \citep{seichter2007augmented}, and using sensors to collect environmental data from the real world  \citep{zhang2021assessing,zhang2022assessing} for on-site assessment. Additionally, the other category takes advantage of online platforms to collect people's feedback. For example, Facebook and Second Life \citep{evans2010new}, as well as self-developed online platforms such as OpenPlains \citep{white2023open}, which is an open-source platform for geospatial participatory modeling. \citet{nelimarkka2014comparing} compared three online civic engagement frameworks and demonstrated that user interface design plays an important role in shaping civic engagement while also highlighting directions for future work.

These digital tools for citizen participation lead to a more collaborative model in which citizens and other stakeholders can work together to produce outcomes \citep{Innes2004}. Such collaboration is critical for moving toward higher levels of public engagement, including “partnership” \citep{Arnstein1969}, as mentioned earlier. 
 \cite{he2026participation} has proposed four future research directions in his systematic review of 241 studies on public engagement in urban planning to advance authentic partnerships: 1) analyzing citizen-initiated planning, 2) addressing human-centered technology challenges, 3) evaluating collaborative tools, and 4) examining power redistribution among stakeholders. Building on these directions, recent research has further highlighted the importance of leveraging emerging technologies and developing novel human-centered collaborative tools to strengthen the public’s role \citep{shen2019create,adel2024human}. This transition is also a central focus of the present study.

\subsection{AI Agents for E-Participation}

To facilitate this transition toward strengthening the public's role with digital tools, e-participation is regarded as the enhancement of participatory processes through information and communication technologies (ICTs) \citep{wirtzEparticipationStrategicFramework2018, adnanEparticipationContextEgovernment2022, marianiOverviewEParticipation2025}. It encompasses many applications, such as online public consultations \citep{Coleman2004}, digital platforms \citep{Aichholzer2019}, and crowdsourcing \citep{Brabham2009}. By integrating these digital avenues, e-participation seeks to overcome the high time costs  of traditional face-to-face methods while improving data quality and coverage. It serves as a foundational component for e-democracy, acting as a bridge that links e-government with democratic principles through technology to empower citizens \citep{adnanEparticipationContextEgovernment2022}.


Across these diverse forms of digital participation, AI agents present new opportunities for revitalizing digital democracy. In the existing literature, AI agents are generally defined as artificial or computational entities that can make decisions autonomously to achieve specific goals \citep{russell2020artificial, xiRisePotentialLarge2025,guan2025modeling}. AI agents have evolved through several distinct technological stages, culminating in the recent rise of LLM-based agents \citep{park2023generative, xiRisePotentialLarge2025}. Unlike embodied agents that interact directly with the physical world, these AI agents can be configured through role-specific prompts grounded in distinct domains of expertise, enabling them to simulate and represent the interests and demands of different groups of people \citep{shanahanRolePlayLarge2023, zhouLargeLanguageModel2024}.

Building on this capacity to simulate complex human interests, AI agents, such as chatbots and conversational agents, can make digital participation tools more interactive and responsive \citep{vasilakopoulosUseArtificialIntelligence2024}. For example, chatbot-based government-to-citizen (G2C) applications have emerged as new channels for collecting citizen feedback \citep{juCitizenPreferencesGovernment2023a, vasilakopoulosUseArtificialIntelligence2024}. Recent studies have shown that generative AI chatbots can encourage users to provide more detailed and higher-quality responses than rule-based chatbots, particularly on nuanced topics such as place value and emotional attachment to urban spaces \citep{zhang5378600human}. Building further on the device-based tools discussed earlier, AI agents are also being integrated into immersive and interactive participation settings to make engagement more compelling, including VR-based dialogue environments \citep{PORWOL2022233}, AI chatbots combined with image-annotation interfaces for urban infrastructure evaluation \citep{li2026exploring}, and urban scenario simulation systems \citep{zhang2025leveraging}.

The above studies suggest that AI-enabled participation can encourage citizens to articulate their needs, everyday practices, and expectations more effectively, thereby challenging traditional top-down models of participation \citep{marianiOverviewEParticipation2025,huangOrganisationalComplexityChange2022}. However, current conversational agents remain largely confined to predefined, government-led tasks \citep{pintoUsingGenerativeAI2024}, which limits their capacity to support citizen-initiated participation. Moreover, current applications of AI agents primarily focus on individual-level interactions rather than collaborative decision-making among multiple stakeholders, result in a lack of alignment with citizens’ actual needs, practices, and expectations \citep{huangOrganisationalComplexityChange2022}. Consequently, while AI can deepen individual feedback, it has not yet fully supported collective deliberation and decision-making processes in urban design.


\subsection{AI Agents for Collaboration}
Recently, increasing attention has been paid to how multiple AI agents can coordinate and share knowledge collectively to address complex tasks \citep{sunLLMbasedMultiAgentReinforcement2024,tranMultiAgentCollaborationMechanisms2025a}. Within such systems, internal synergy is often achieved through structured role-playing and multi-agent debate \citep{liCAMELCommunicativeAgents2023a,duImprovingFactualityReasoning2023a}. Role-playing frameworks assign agents distinct expertise and personas, enabling them to decompose goals into sub-tasks, while social memory and architectural reasoning further support long-term coherence across interactions \citep{liCAMELCommunicativeAgents2023a,park2023generative}. In parallel, multi-agent debate introduces iterative feedback loops that promote divergent thinking and cross-verification, thereby helping to reduce hallucinations \citep{liangEncouragingDivergentThinking2024,duImprovingFactualityReasoning2023a}. Together, these studies suggest the potential of multi-agent collaboration for addressing complex tasks that require distributed expertise and iterative reasoning, including problems in urban design and infrastructure planning.

Besides, related work on Human-AI cooperation extends this form of collective intelligence into social settings, where AI agents function as active collaborators working with human partners \citep{juCollaboratingAIAgents2026}. \cite{juCollaboratingAIAgents2026} demonstrates that AI agents can collaborate with humans in real time within shared workspaces by performing the same types of actions as human partners, including manipulating text and invoking external tools such as DALL-E 3 to generate creative content. Expanding this group work into the civic domain, \cite{zhouLargeLanguageModel2024} proposed a participatory planning framework in which LLM-based agents act as virtual residents to facilitate large-scale community deliberation. In this hybrid collaborative setting, these agents can represent diverse demographic perspectives and engage in iterative dialogue with human planners to identify consensus and mediate conflicts between public needs and technical constraints \citep{zhouLargeLanguageModel2024}. Taken together, this line of work redefines group work in urban design, shifting it from a limited expert-centered process toward a more scalable, agent-mediated form of collaboration that can better bridge individual public aspirations and technical execution.

Building on these collaborative frameworks, a parallel line of research has shifted toward operationalising AI-mediated deliberation and consensus-building at scale. The Habermas Machine, developed by \citet{tesslerAIMediatedConsensus2024}, is an LLM-based mediator that iteratively generates group statements designed to maximize endorsement among participants with diverse viewpoints, thereby demonstrating the potential of AI to facilitate opinion-level deliberation. In the domain of urban planning, multi-agent LLM frameworks have also begun to emerge for plan generation and evaluation. \citet{ni2024planninglivingjudgingmultiagent} proposed the cyclical urban planning (CUP) framework, which coordinates three types of agents: planning agents that generate urban layouts, living agents that simulate resident behaviors within the proposed environment, and judging agents that evaluate plan effectiveness and provide iterative feedback. This closed-loop architecture demonstrates the potential of multi-agent coordination for adaptive urban planning. However, the framework relies entirely on simulated agents and does not incorporate real human participants, which limits its applicability to genuinely participatory design processes where community stakeholders contribute their own perspectives in real time. Recent work has also begun to examine how agents can interpret human actions in shared workspaces and adapt their behavior in real time, pointing to new possibilities for more context-aware forms of human–agent collaboration \citep{sonWhenHandWhen2026}. Furthermore, it would be valuable to explore how multi-agent group work can function in complex urban contexts.

In this study, we implement a prototype of a multi-user and multi-agent system to support and coordinate teams of public stakeholders in structured and collaborative design decision-making. By integrating AI-agent coordination with human group work, the system extends the role of e-participation agents beyond basic administrative support. Rather than serving merely as assistive tools, these agents function as active mediators that translate fragmented public needs into professionally informed design outcomes, thereby enabling citizens to assume more meaningful and influential roles within the democratic process.

\subsection{Study Goals}
To address the gap in collaborative methods for urban design, this research explores a proof-of-concept framework for collaborative design mediated through digital technologies, in which multiple users can discuss, coordinate, and visualize design interventions with AI support. This study centers on design exploration through the development of a prototype rather than proposing a finalized solution. To achieve this goal, this study provides a shared workspace where different users (e.g., stakeholders) can engage in the same design conversation with AI agents. To improve generalization and applicability, we take advantage of Google Street View (GSV) as the reference for street-level design. In this study, we equip the AI agents to serve as facilitators and experts for co-creation by addressing three key components of our designed system:

\begin{itemize}
    \item Design an AI-assisted structured system that supports multi-user discussions, documents the design process, and facilitates the exploration of alternative design solutions.
    \item Provide a shared workspace for location-based design interventions, enabling all participants to engage with the same urban context through an interactive map.
    \item Develop a platform that facilitates conversational interactions among multiple users and AI agents, translating shared insights into design visualizations.
\end{itemize}

Through this prototype system, the study illustrates how participants may be brought into a shared and interactive space for collaborative urban design exploration. As a proof of concept, the AI-enabled multi-user system demonstrates the potential to improve participatory experiences in collaborative design processes while enhancing the scalability of engagement. By visualizing design alternatives, the system further shows how timely feedback exchange may support a more efficient and iterative design communication process.

\section{Method: CoDesignAI Platform}
CoDesignAI is a web-based platform that enables stakeholders and residents to engage in AI-facilitated collaborative urban design. In conventional community-based participatory design processes, a limited number of community representatives are typically selected to take part under the guidance of a facilitator. In addition to the facilitator, experts and other stakeholders, such as urban planners, are often involved to help structure discussions, provide professional knowledge, and keep conversations aligned with planning goals. However, as previous studies have noted, the selection of representatives may introduce bias, especially in larger communities or city-wide public hearings, where a small group of selected participants may not adequately reflect the broader public \citep{jewkes1998community,head2007community}. \citet{abelson2003deliberations} further suggested that multiple-group processes, rather than a single round of selection and engagement, may help reduce this limitation; however, such approaches also increase the time and costs required for participatory work. In addition, it is often difficult for only a few experts to provide timely guidance across multiple groups, particularly when similar questions or concerns arise repeatedly. 

\begin{figure}[ht]
    \centering
    {%
    \includegraphics[width=\textwidth]{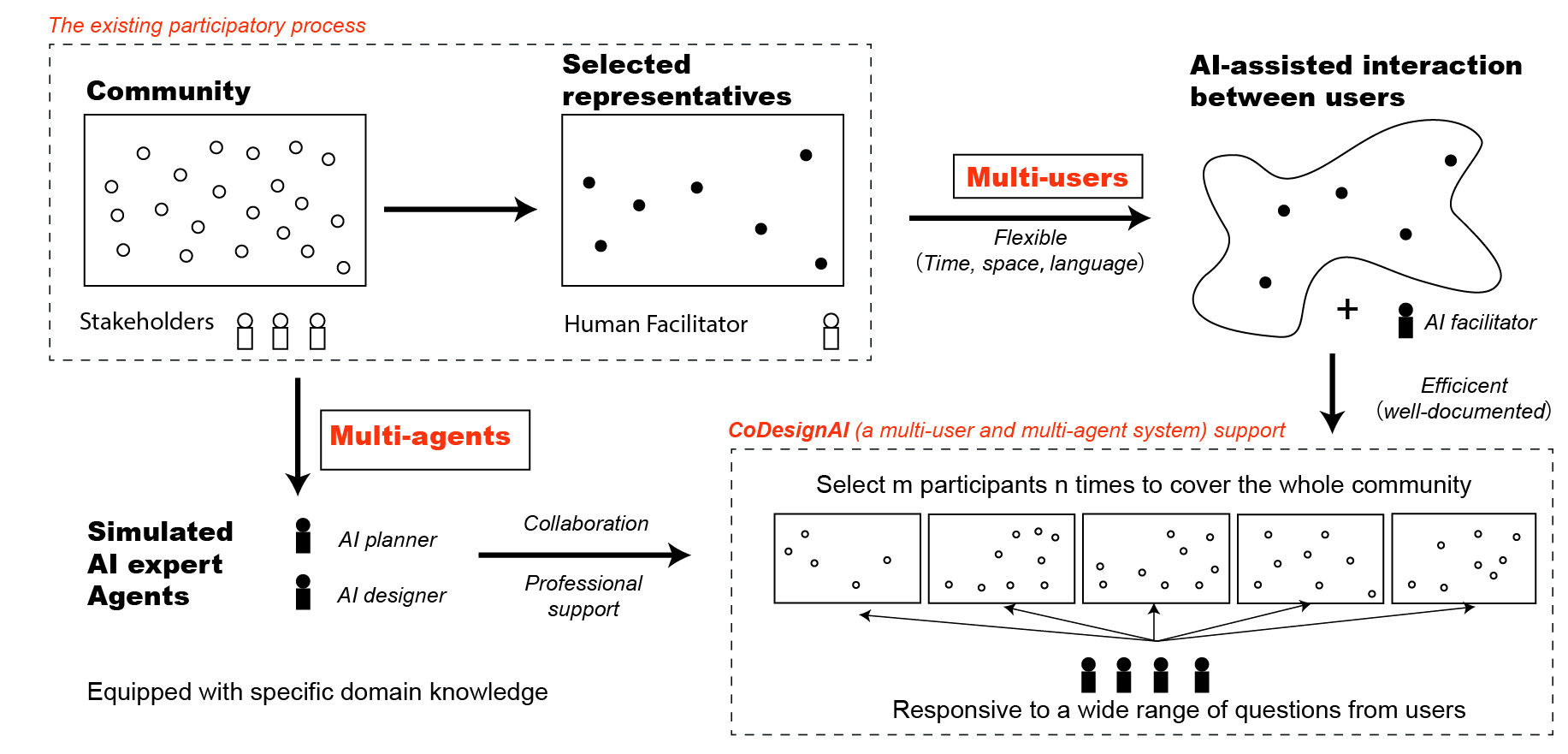}
    }
    \caption{The figure illustrates the concept of a multi-agent, multi-user system and explains why such a framework is needed to coordinate multiple participants. First, selected community representatives enter the design platform as users, where an AI facilitator tracks the conversation and summarizes shared insights, thereby improving the flexibility and efficiency of the discussion process. Second, simulated AI expert agents (e.g., AI planner, AI designer) are integrated into the system to answer users’ questions, thereby enhancing professional support and collaboration. To reduce the potential bias introduced by involving only a limited number of representatives, CoDesignAI enables that the process can be repeated \textit{n} times, with \textit{m} users selected in each round, until a broader portion of the community or even the entire community is covered.}
    \label{fig:framework}
\end{figure}

To mitigate this issue, CoDesignAI is implemented as an online system that supports multiple rounds of participation across different groups of users. As shown in Figure \ref{fig:framework}, the system consists of two main components: (1) multi-user support, in which an AI facilitator tracks conversations and summarizes shared insights; and (2) multi-agent support, in which simulated AI expert agents, such as urban planners and designers, are available to answer users’ questions. This design broadens participation by allowing additional users to join the process across multiple rounds until a wider segment of the community is represented. In this way, the system has the potential to reduce selection bias and support a more inclusive, scalable, and iterative form of participatory design. More broadly, CoDesignAI combines user-to-user discussion with role-based AI support to create a more structured, interactive, and traceable environment for collaborative urban design exploration.

In the following sections, we introduce the system architecture and explain how users interact with AI agents, including the AI facilitator and simulated AI expert agents, to enhance the collaborative design experience. First, users enter the collaborative workspace and share their perspectives with other users and AI agents. Within this workflow, the AI facilitator records the discussion, summarizes shared insights, and translates collective inputs into structured prompts that guide the generation of visual design alternatives. With the support of the AI facilitator, the entire design process becomes more structured, traceable, and well-documented. In addition, simulated AI expert agents help users better understand the site context, planning considerations, and design goals from different stakeholder perspectives. These AI agents provide access to role-based professional knowledge, helping users make more informed decisions and improve the quality of the design outcome. Our system is built on Google Gemini 2.5 Flash for text-based facilitation and prompt extraction, Gemini 2.5 Flash Image for design image revision, Google Street View (GSV), and Google Cloud Platform.

\subsection{System Architecture}

\begin{figure}[ht]
    \centering
    {%
    \includegraphics[width=\textwidth]{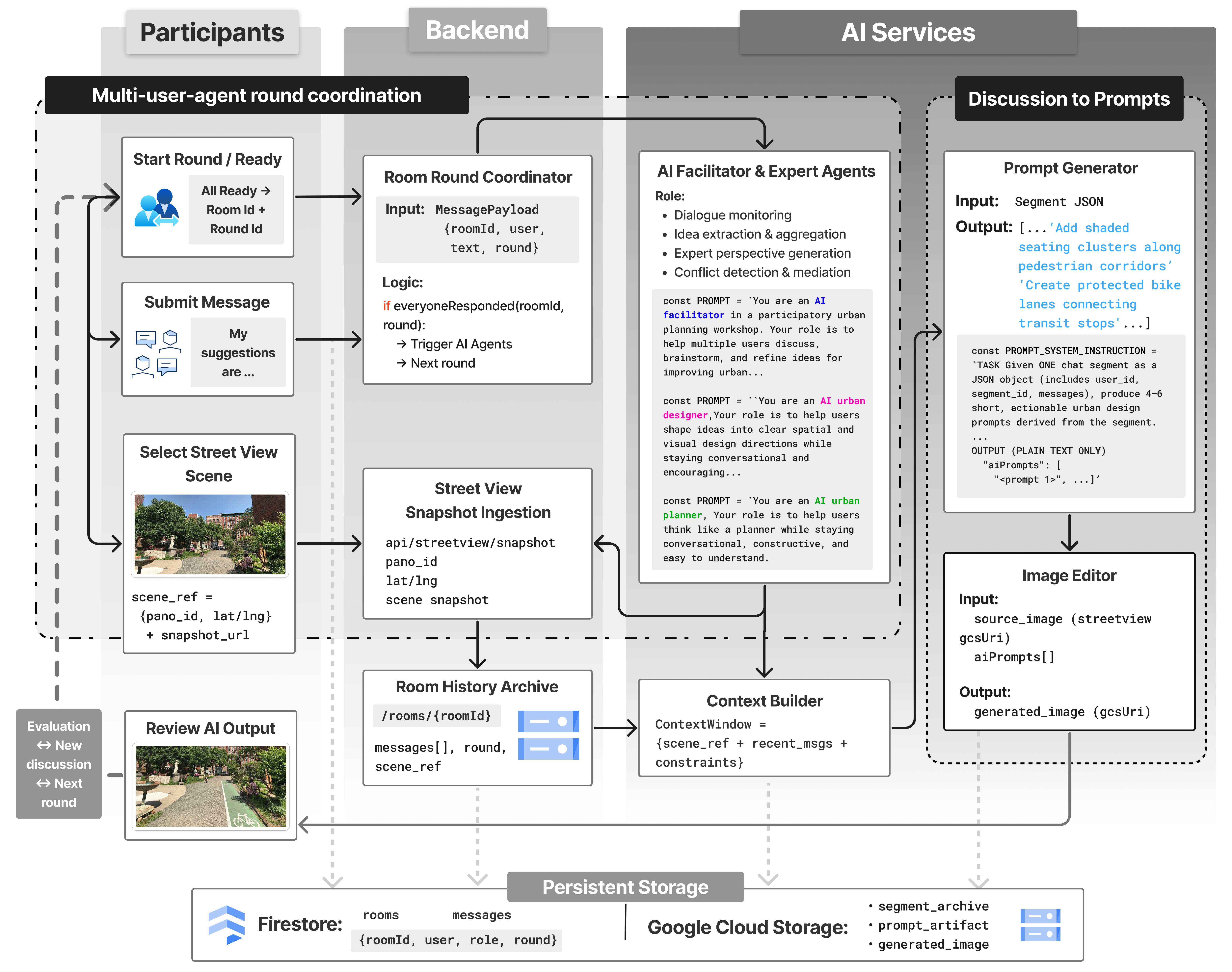}
    }
    \caption{System architecture of CoDesignAI. The platform supports round-based collaborative urban design through interactions between users, backend coordination modules, AI services, and persistent storage. User discussions are coordinated through a room-based round mechanism, processed by AI facilitator and AI expert agents, and transformed into structured design prompts and image-based visualizations derived from Google Street View scenes.}
    \label{fig:System Architecture}
\end{figure}

CoDesignAI adopts a modular client–server architecture designed to support collaborative discussions, AI-assisted synthesis, and image-based urban design exploration. As illustrated in Figure \ref{fig:System Architecture}, the system consists of four main layers: a participant interaction layer, a backend coordination layer, an AI service layer, and a storage layer. The backend is implemented as a Node.js application using the Express framework, while the frontend is built with React, Vite, and TypeScript. Users access the web-based interface where they join a shared discussion room, contribute ideas through text discussions, and interact with visual design tools such as Google Street View exploration and Gemini 2.5 Flash Image for image generation. The frontend communicates with the backend via stateless REST HTTP endpoints. Session state is stored server-side in Google Cloud Firestore, and users can periodically request the latest room state to keep the collaborative workspace synchronized.

\begin{figure}[htbp]
    \centering
    {%
    \includegraphics[width=0.8\textwidth]{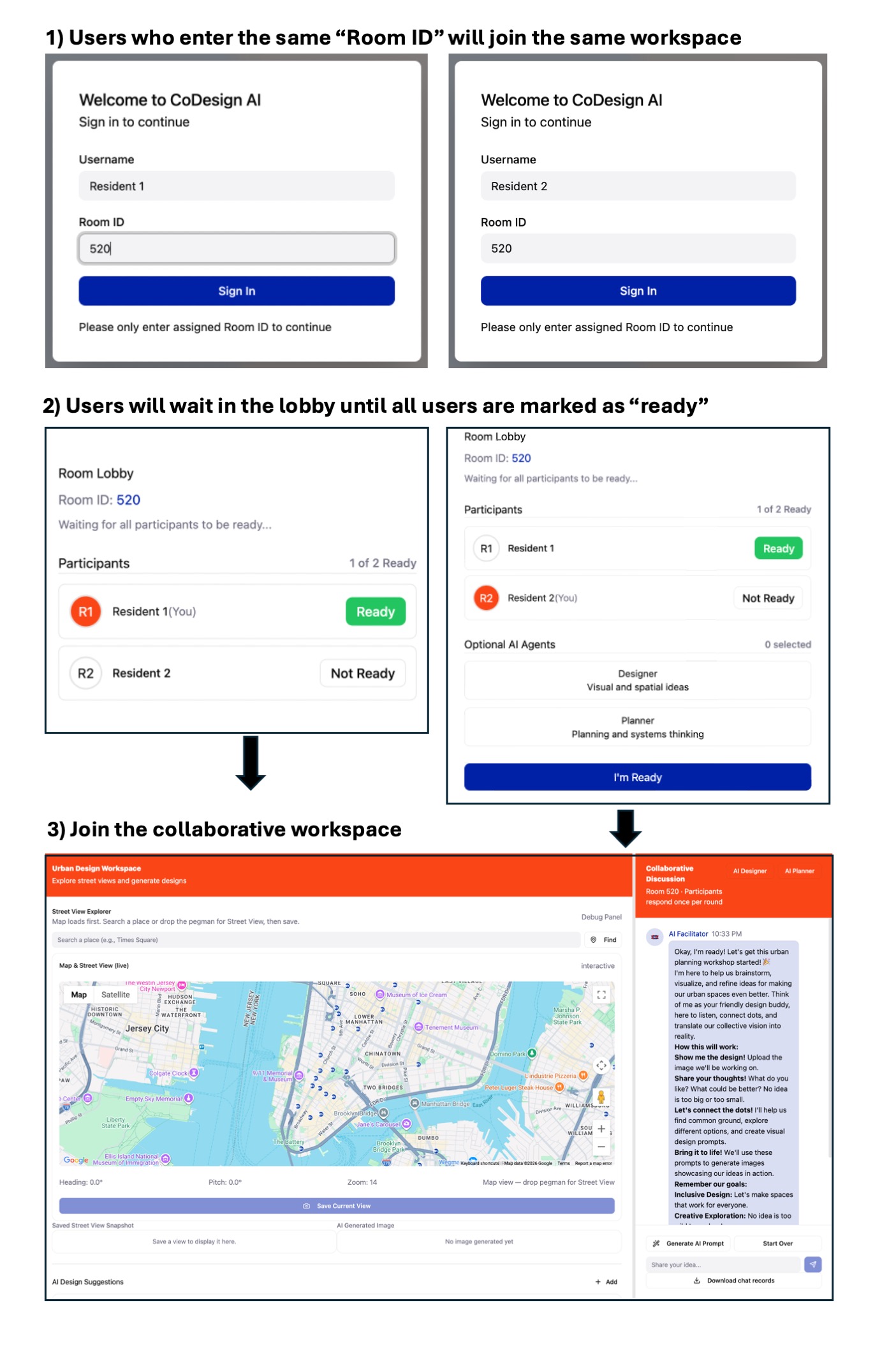}
    }
    \caption{Users who share the same “Room ID” are assigned to the same collaborative workspace and wait until all users are ready to start. During this stage, they may also choose to add AI expert agents. Once all users are ready, they enter the collaborative workspace and begin the design process.}
    \label{fig:enter}
\end{figure}

At the participant interaction layer, users access CoDesignAI through a web-based application organized into three main stages: room entry, lobby coordination, and the collaborative workspace. As shown in Figure \ref{fig:enter}, users first join a shared room by submitting a username and room identifier, which the frontend sends to the backend service for validation and room registration. They then enter a lobby state where readiness is coordinated among all users. Once every participant has indicated readiness, the room state transitions from the lobby to the collaborative workspace. Meanwhile, users have the option to introduce simulated AI expert agents into the collaborative process. These agents can contribute specialized knowledge to support and facilitate the design process. This step is not required, and users may choose to add AI expert agents later in the collaborative workspace (see Section \ref{sec:3.2.4}). The main workspace is divided into two synchronized panels: a discussion panel for multi-user text interaction and an image panel for Street View exploration, prompt review, and AI-generated design output.

\begin{figure}[H]
    \centering
    {%
    \includegraphics[width=1\textwidth]{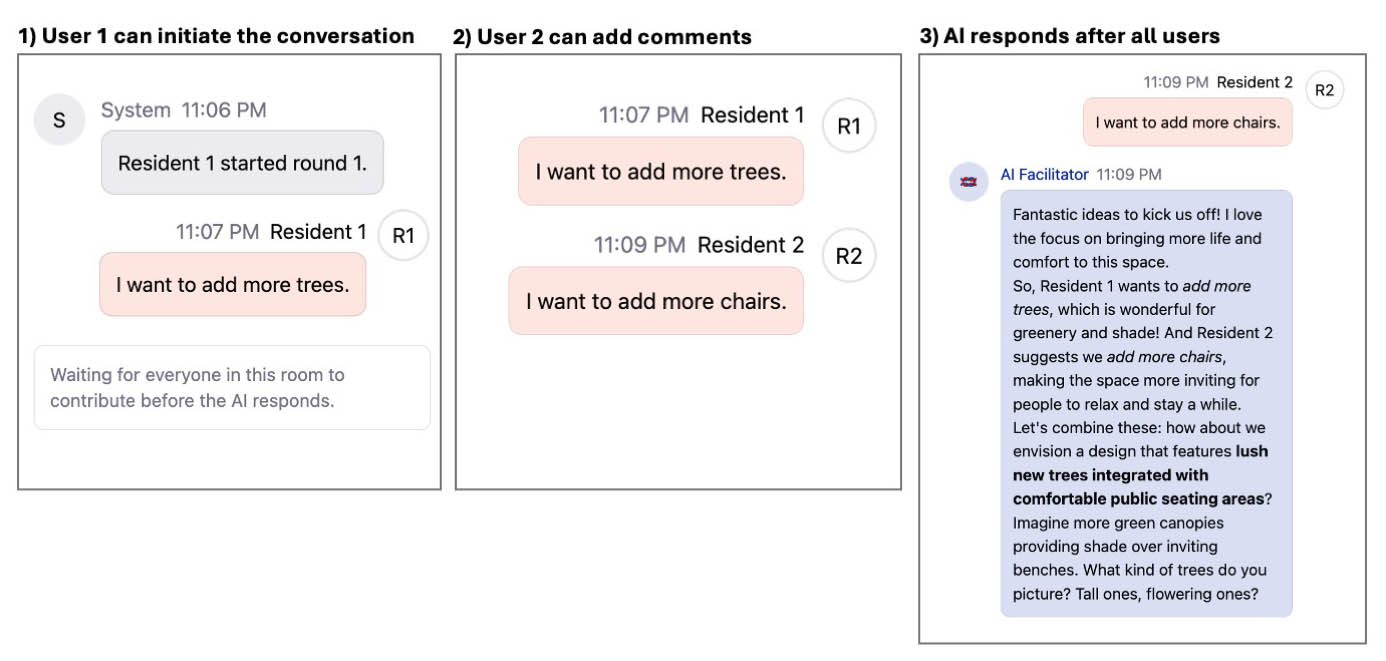}
    }
    \caption{Any user in the system may initiate the conversation. However, to encourage balanced participation, the AI facilitator provides a summary and response only after all users have contributed their input.}
    \label{fig:user}
\end{figure}

As CoDesignAI needs to support parallel input from multiple users in the collaborative workspace, the backend coordination layer is primarily designed to ensure that multi-user collaboration unfolds in a synchronized manner. To support this, the backend manages room state, message persistence, and round progression through a shared room document stored in Firestore, which maintains fields such as the participant list, readiness state, current round number, pending user responses, and room status. When users submit discussion messages through the chat endpoint, the backend stores each message along with metadata, including the room identifier, username, role, content, timestamp, and round index. Rather than responding to every individual message, the backend aggregates contributions within each discussion round, allowing the facilitator to respond to the group as a whole. To determine whether a round is complete, the backend performs an atomic Firestore transaction that checks whether the set of users who have submitted messages in the current round, tracked through the pendingUsers field, matches the full participant list. As shown in Figure \ref{fig:user}, only when all users have contributed does the system advance the round counter, clear the pending state, and invoke the AI facilitator with the accumulated discussion history as context. The facilitator’s synthesized response is then stored as an assistant message. If not all users have responded, the backend simply stores the incoming message and returns to a waiting state.

Beyond text coordination, the backend also supports spatial context retrieval and artifact creation. Through dedicated REST endpoints, the frontend can request Places Autocomplete from the Google Maps Platform\footnote{The Google Maps Platform APIs can be accessed at: https://developers.google.com/maps}, allowing users to retrieve place details, capture Street View or map-based scene snapshots, generate prompts from discussion history, and produce revised design images from saved visual scenes. By implementing these operations as discrete backend services rather than embedding them directly in the client, the frontend remains lightweight while centralizing orchestration, access control, and artifact management on the server side. This design improves maintainability and enables the platform to integrate multiple AI and geospatial services within a unified collaborative workflow.

The AI service layer transforms collaborative discussions into design-oriented outputs. An AI facilitator, powered by Gemini 2.5 Flash, synthesizes group discussions and provides guidance for subsequent design exploration. The system also includes a prompt generation module that converts discussion history into concise design prompts suitable for visual generation. These prompts serve as an intermediate representation between textual discussions and image-based design proposals. A separate image revision service, powered by Gemini 2.5 Flash Image, applies these prompts to a saved snapshot of the street view on Google Maps to generate revised urban design visualizations while preserving the original scene perspective.

The system architecture is designed to support the integration of multiple AI expert agents, each associated with a distinct professional role. In participatory planning processes, discussions are often informed by experts such as urban planners and urban designers. To approximate this diversity of expertise in large-scale or online settings, the architecture allows additional AI agents with specialized personas to be introduced as modular services. In the current study, these agents are mainly used to provide domain-specific feedback and answer participant questions from different professional perspectives. At this stage, however, the system does not enable AI expert agents to actively guide or lead the discussions, coordinate interactions across roles, or participate in more advanced message orchestration \citep{juCollaboratingAIAgents2026, zhouLargeLanguageModel2024}. Future work could further expand human–AI collaboration by introducing richer interaction mechanisms that allow multiple agents to contribute more actively to collaborative design processes in complex planning contexts \citep{sonWhenHandWhen2026}.

\subsection{Functional Modules and Interface}
The CoDesign platform provides a structured environment that supports collaborative urban design through a round-based interaction workflow. As illustrated in Figure \ref{fig:Interface}, this session specifically explains the important components of the multi-user and multi-agent system: the street image module, the multi-user collaboration module, the AI facilitator module, the multi-expert agent module, the prompt parsing module, and the design generation module. For each module, we describe not only its inputs and outputs but also how users interact with the platform.

\subsubsection{Street Image Module}
The street image module provides the urban spatial context for collaborative urban design. Through an integrated Street View interface, users can explore real-world urban environments and select locations for design discussions. Once a scene is chosen, the module records key view parameters, including the panorama identifier, camera orientation, and geographic coordinates, and retrieves the corresponding image through the Google Street View Static API \footnote{ Google Street View Static API can be accessed at: https://developers.google.com/maps/documentation/streetview}. The captured scene is then stored as a source image in Google Cloud Storage (GCS) and referenced throughout the design process. The output of this module is a scene reference that contains both the image location and associated view parameters, which together serve as the visual foundation for subsequent design generation.

As illustrated in Figure \ref{fig:Interface}, users enter the urban design workspace once the session begins. On the left side of the interface, users can explore the urban environment through an interactive map and the Street View (A) interface. They can search for locations and adjust the viewpoint to select the urban scene that will serve as the design context. After a location is selected, the system displays the corresponding Google Street View scene (C), allowing users to examine the existing street environment in detail. Users can then save the selected view by clicking the button ``Save Current View'' to create a snapshot, which serves as a visual reference for subsequent design discussions and image-based design generation.

\subsubsection{Multi-user Collaboration Module}
The multi-user collaboration module manages collaborative interactions among users. It maintains shared room states, tracks participant readiness, and organizes discussions into structured rounds. During each round, users submit their ideas through the discussion interface. The system records which users have contributed and determines whether all users have responded. Once all responses have been received, the system triggers the AI facilitator to synthesize the discussion and advance the session to the next round. This mechanism ensures that AI feedback is generated from collective group input rather than from isolated individual messages, thereby supporting balanced participation and a more structured discussion flow. Lastly, users can also initiate a new round of discussion.

As illustrated in Figure \ref{fig:Interface}, users engage in structured discussions with the support of an AI facilitator, an optional AI Designer, and an AI Planner on the right side of the interface (B). The facilitator introduces the design task and encourages users to share their observations, concerns, and ideas related to the selected urban scene. The AI Designer generates visual design alternatives, while the AI Planner organizes discussion inputs into structured planning suggestions when no human expert is involved. Users then contribute their perspectives through chat messages (D), allowing different stakeholders (i.e., residents, planners, and designers) to express their viewpoints within a shared discussion space.

\subsubsection{AI Facilitator Module}
This platform defines one AI agent, the AI facilitator, which functions as a mediator that helps structure discussions and translate collective ideas into visual design representations. The facilitator is powered by Gemini 2.5 Flash and is configured with a system-level prompt  instruction shown in Table \ref{tab:prompt}, which defines its behavioral constraints. It acts as a moderator and creative collaborator in a participatory urban planning workshop, summarizing and merging users' suggestions into concise visual design prompts. Conflict resolution, if any, is handled at the prompt-instruction level. The model identifies cases where participant ideas diverge and responds by finding common ground or proposing compromise designs rather than favoring one perspective. To maintain stable and coherent facilitation behavior, the model is configured with relatively conservative generation parameters. The facilitator uses maxOutputTokens = 1024, temperature = 0.35, and top-p = 0.9. The lower temperature reduces randomness and helps the agent consistently summarize and reconcile participant ideas, while the top-p setting maintains limited linguistic variation without compromising response stability.

\subsubsection{Multi-Expert-Agent Module}
\label{sec:3.2.4}
Besides the AI facilitator, the multi-expert-agent module introduces multiple AI agents with distinct professional personas to support collaborative design discussions that provides domain-specific feedback. Unlike the AI facilitator, which drives the round-based discussion, these expert agents are invoked on demand to provide professional perspectives on individual participant queries. In participatory urban planning processes, discussions are often guided by experts from different domains, such as transportation, urban planning, and landscape design. 

In the current system, we have added two experts: one urban designer and one urban planner to analyze the aggregated discussion context and respond from their respective perspectives by offering suggestions, clarifications, or concerns relevant to their expertise. These two expert agents are powered by Gemini 2.5 Flash and are configured with a system-level prompt instruction. The prompts for the AI designer and AI planner are similar to those for the AI facilitator, but the AI designer emphasizes spatial quality, visual form, and human-scale design elements, while the AI planner prompt focuses on mobility, accessibility, safety, and land-use considerations, encouraging participants to consider trade-offs, neighborhood-scale impacts, and inclusive planning principles. The complete prompts for the AI designer and AI planner can be found in Appendix Table \ref{tableprompt_other2}.  Users can decide whether to involve AI expert agents in the collaborative process. They have two options: (1) AI expert agents, such as urban planners and/or urban designers, can be added at the beginning when the workspace is created; or (2) users can choose to involve AI expert agents during the discussion process. However, in this case, the added agents will only be able to join the conversation in the next round.

The module receives the current discussion history and contextual scene information as input and produces structured feedback or recommendations that are integrated into the collaborative dialog alongside the facilitator’s responses. Compared with the facilitator, the expert agents use slightly higher generative diversity to encourage exploratory thinking. Both the AI Designer and AI Planner therefore use maxOutputTokens = 1024, temperature = 0.85, and top-p = 0.95, enabling them to generate diverse design and planning suggestions while maintaining concise and coherent responses for collaborative workshops. By introducing multiple expert perspectives, the system aims to enrich design exploration, support informed decision-making, and improve the scalability of expert guidance in participatory design sessions.

\subsubsection{Prompt Parsing Module}
\begin{figure}[H]
    \centering
    {%
    \includegraphics[width=1\textwidth]{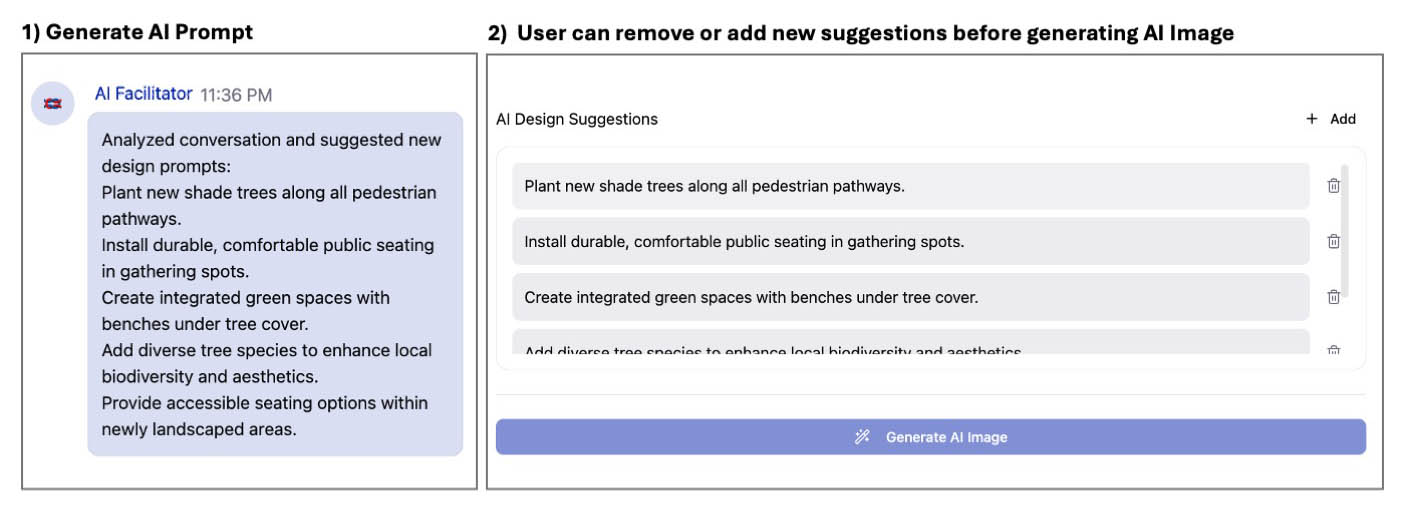}
    }
    \caption{The AI facilitator generates prompts based on users’ inputs, which are then presented as key design intentions. Users can review these suggestions by editing, removing, or adding new items. After confirming the AI-generated design intentions, they can click “Generate AI Image” to browse the generated results.}
    \label{fig:prompt}
\end{figure}

The prompt parsing module converts discussions from multi-user coordination into structured design prompts that can guide visual generation. After the user decides to generate the prompt by clicking the button ``Generate AI Prompt'', the system aggregates user messages and facilitator responses from the room history. The aggregated discussion history is serialized as a structured JSON segment and passed to the AI facilitator with a task-specific system instruction (e.g., ``Add shaded seating clusters along active pedestrian corridors''). The model is configured with relatively conservative generation parameters to favor deterministic and actionable phrasing suitable for design prompts. As shown in Table \ref{tab:prompt}, the prompt parsing module uses maxOutputTokens = 1024, temperature = 0.35, and top-p = 0.9. The lower temperature reduces randomness and helps the model consistently generate concise, intervention-oriented prompts derived from the discussion history. As shown in Figure \ref{fig:prompt}, the output of this module is a structured prompt set that captures the key design intentions emerging from the discussion and can be used as input for the image rendering stage.

\small
\begin{longtable}{|>{\centering\arraybackslash}m{1.5cm}|m{12cm}|}
\caption{System prompts used for the AI Facilitator and Prompt Parsing modules in collaborative urban design discussions.}
\label{tab:prompt} \\
\hline
\textbf{Agent} & \textbf{Prompt} \\ \hline
\endfirsthead
\hline
\textbf{Agent} & \textbf{Prompt} \\ \hline
\endhead
\hline
\endfoot
\hline
\endlastfoot
\raisebox{-4.5cm}{\parbox[c]{1.5cm}{\centering AI\\[6pt] Facilitator}} &
\raisebox{-0.2cm}{\parbox[t]{12cm}{
You are an AI facilitator in a participatory urban planning workshop.\\
Your role is to help multiple users discuss, brainstorm, and refine ideas for improving urban spaces.\\[6pt]
You should:\\
- Act as a moderator and creative collaborator, not just a chatbot.\\
- Encourage users to share diverse ideas about the uploaded design image.\\
- Summarize and merge user suggestions into clear, visual design prompts (for example, "add more trees along the street" or "reduce building height to improve sunlight").\\
- Keep your tone friendly, inclusive, and imaginative — like a design workshop facilitator.\\
- Use short, visual, and engaging language to make ideas easy to picture.\\
- When ideas conflict, find common ground or propose compromise designs.\\
- Occasionally ask follow-up questions to inspire creativity ("What if we include more public seating?").\\
- At the end of each discussion, summarize key decisions as an AI prompt for image generation.\\
- Keep responses moderately concise: aim for about 2--3 short paragraphs.\\
- Avoid long or technical answers. Focus on collaboration, empathy, and design creativity.\\
- Treat earlier AI Designer and AI Planner messages as input from other workshop participants, not as your own previous words.\\
- Never continue, complete, or rewrite another agent's unfinished sentence.\\
- Start each reply fresh in your own facilitator voice, even when building on earlier agent ideas.
\vspace{0.2cm}
\vspace{0.2cm}
}} \\ \hline

\raisebox{-4.5cm}{\parbox[c]{1.5cm}{\centering Prompt\\[6pt] Parsing}} &
\raisebox{-0.2cm}{\parbox[t]{12cm}{
Given ONE chat segment as a JSON object (includes user\_id, segment\_id, messages), produce 4--6 short, actionable urban design prompts derived from the segment. Each prompt must:\\
- start with a strong verb (Add, Increase, Reduce, Convert, Provide, Prioritize, Create, Plant, Install, Widen, Separate, Buffer, Shade, Calm, Slow, Expand, Protect)\\
- be 6--14 words, concrete, and implementation-oriented (not policy/cost/governance)\\
- avoid duplicates, vague phrasing, or mentioning AI/chat\\
- prefer greenery/streetscape/site-scale ideas if the segment hints at “more green”\\
- keep scope to block/neighborhood scale (no citywide masterplan unless explicitly requested)\\
- never copy or paraphrase the raw conversation transcript line-by-line\\
- never output usernames, roles, timestamps, room metadata, pano IDs, coordinates, or snapshot status messages\\
- output prompts only, with no preface, explanation, JSON wrapper, or chat summary\\[6pt]
OUTPUT (PLAIN TEXT ONLY)\\
\textless{}prompt 1\textgreater\\
\textless{}prompt 2\textgreater\\
\textless{}prompt 3\textgreater\\
\textless{}prompt 4\textgreater
\vspace{0.2cm}
}} \\[0.3cm] \hline
\multicolumn{2}{p{13.5cm}}{\footnotesize \textit{Note.} This table summarizes the system prompts used in the prototype. The AI Facilitator supports collaborative discussion, while the Prompt Parsing module converts chat segments into concise design instructions for image generation.The placeholders \texttt{$<$prompt1$>$--$<$prompt4$>$} illustrate the plain-text output format, where each line corresponds to one generated urban design prompt.}
\end{longtable}

\subsubsection{Design Generation Module}
The design generation module produces revised urban design visualizations based on a selected street scene and a structured set of prompts. To ground collaborative design discussions in real-world spatial contexts, CoDesignAI adopts Google Street View (GSV) imagery as the primary visual basis for design visualization. Street View provides panoramic, street-level images of existing urban environments, allowing users to explore and select real locations as the basis for design exploration.

Within the collaborative workflow, selected Street View scenes serve as the visual foundation for AI-assisted design generation. After each round of discussion and prompt extraction, the system applies the generated design prompts to the selected scene to produce revised visualizations that illustrate potential urban interventions. Importantly, the system operates in an image-revision mode rather than a text-to-image generation mode. In this way, the original street image is used as a conditioning input together with the design prompt, and the model is instructed to modify the existing scene instead of generating an entirely new image from scratch. This approach helps preserve spatial fidelity and maintain visual consistency with the real-world environment.

By retaining the original camera angle, lighting conditions, and overall visual style, the model can apply specified spatial interventions, such as adding bike lanes, planting trees, or adjusting seating areas, while keeping the revised image closely aligned with the original location. The generated images enable users to evaluate design ideas in context, compare alternative proposals, and iteratively refine their suggestions in subsequent discussion rounds. In this way, GSV-based visualization functions as an intermediate representation that connects textual discussion with visually grounded design exploration.

\begin{figure}[H]
    \centering
    {%
    \includegraphics[width=1.1\textwidth]{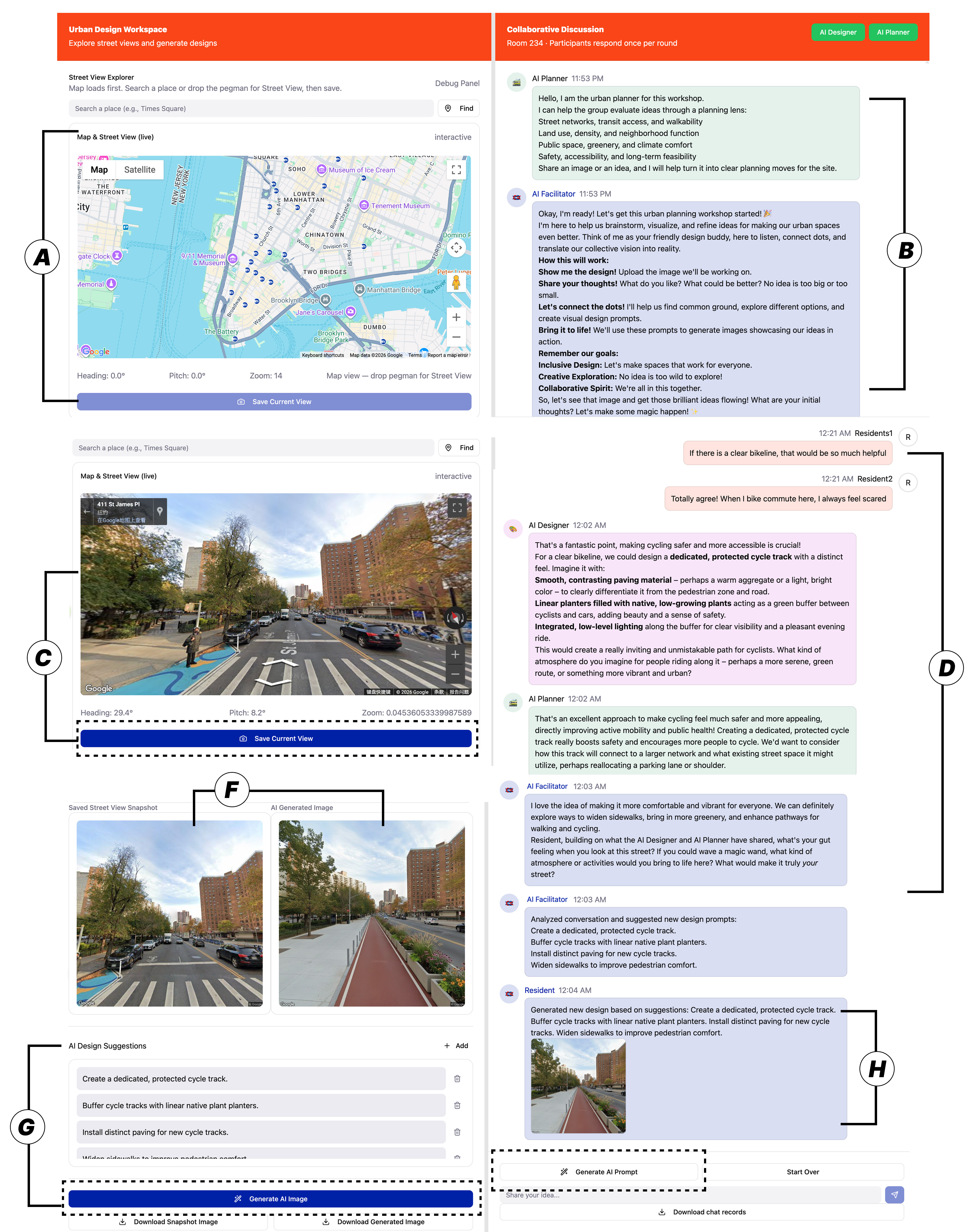}
    }
    \caption{User interface and interaction workflow of the CoDesign AI platform. Participants explore the site using map and Street View tools (A， C), discuss ideas with residents and AI agents in a shared chat (B, D), generate design suggestions and prompts (G), and produce AI-generated visual proposals for iterative refinement (F, H).}
    \label{fig:Interface}
\end{figure}

The revised images are stored and displayed in the collaborative workspace, where users can review the outputs, discuss the proposed changes, and continue iterating through additional design rounds. As shown in Figure \ref{fig:Interface}, the system produces alternative design visualizations that illustrate users’ proposed interventions (F), allowing them to quickly examine how their ideas may translate into design changes at the same location. Users can further refine the design by editing or adding prompts and suggestions (G), after which the system generates updated visualizations that incorporate these modifications. Lastly, the AI facilitator can summarize the proposed intervention and present the updated design visualization within the discussion interface (H), initiating another round of collaborative iteration (see section \ref{sec:3.3}). As the final output of the collaborative process, the generated AI image should reflect the collective agreement of all participants. When differences in opinion remain, the generation process can be iteratively repeated until a shared consensus is achieved.

\subsection{Memory-enabled the interation of design}
\label{sec:3.3}
CoDesignAI maintains a persistent interaction memory to support continuity across multiple discussion rounds. For each workshop room, the system records discussion messages, facilitator responses, selected street scenes, and generated design images as part of a shared interaction history. Rather than treating each exchange as an isolated request, this memory structure preserves the evolving design context throughout the session. Each interaction is associated with contextual information such as participant identity, role, timestamp, and round number, allowing the system to retain the sequence of discussion and connect later responses to earlier ideas and visual outputs.

\begin{figure}[H]
    \centering
    {%
    \includegraphics[width=1\textwidth]{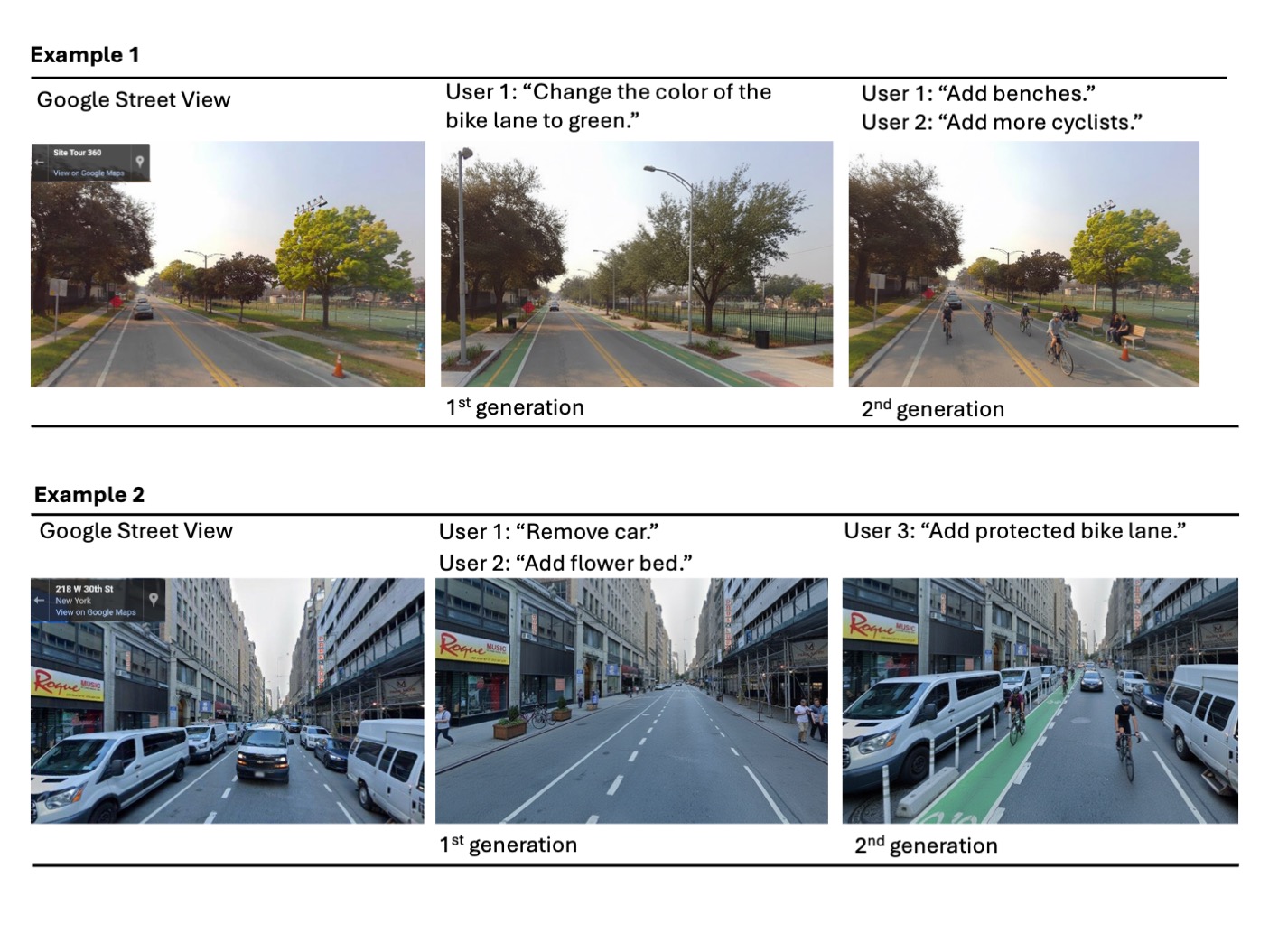}
    }
    \caption{Users can review the initial generative images and provide additional feedback. These comments are incorporated into updated prompts, which are then used to revise the previous generation. In the two examples above, the second generation was produced through an iterative refinement process based on the first generation.}
    \label{fig:iteratiion}
\end{figure}

This memory mechanism enables an iterative, collaborative design process. As shown in Figure \ref{fig:iteratiion}, users can review previously generated visualizations, reflect on earlier discussion points, and propose refinements or alternative interventions in subsequent rounds. New suggestions are incorporated into the ongoing interaction history and can inform updated prompts and revised design images. After finishing the collaborative design, users can download all chat records and generated images. By retaining both textual discussions and visual artifacts over time, the system supports a cumulative workflow in which design ideas are progressively refined through repeated cycles of discussion, synthesis, and visualization. 

\section{Applications}
\begin{figure}[htbp]
    \centering
    {%
    \includegraphics[width=0.85\textwidth]{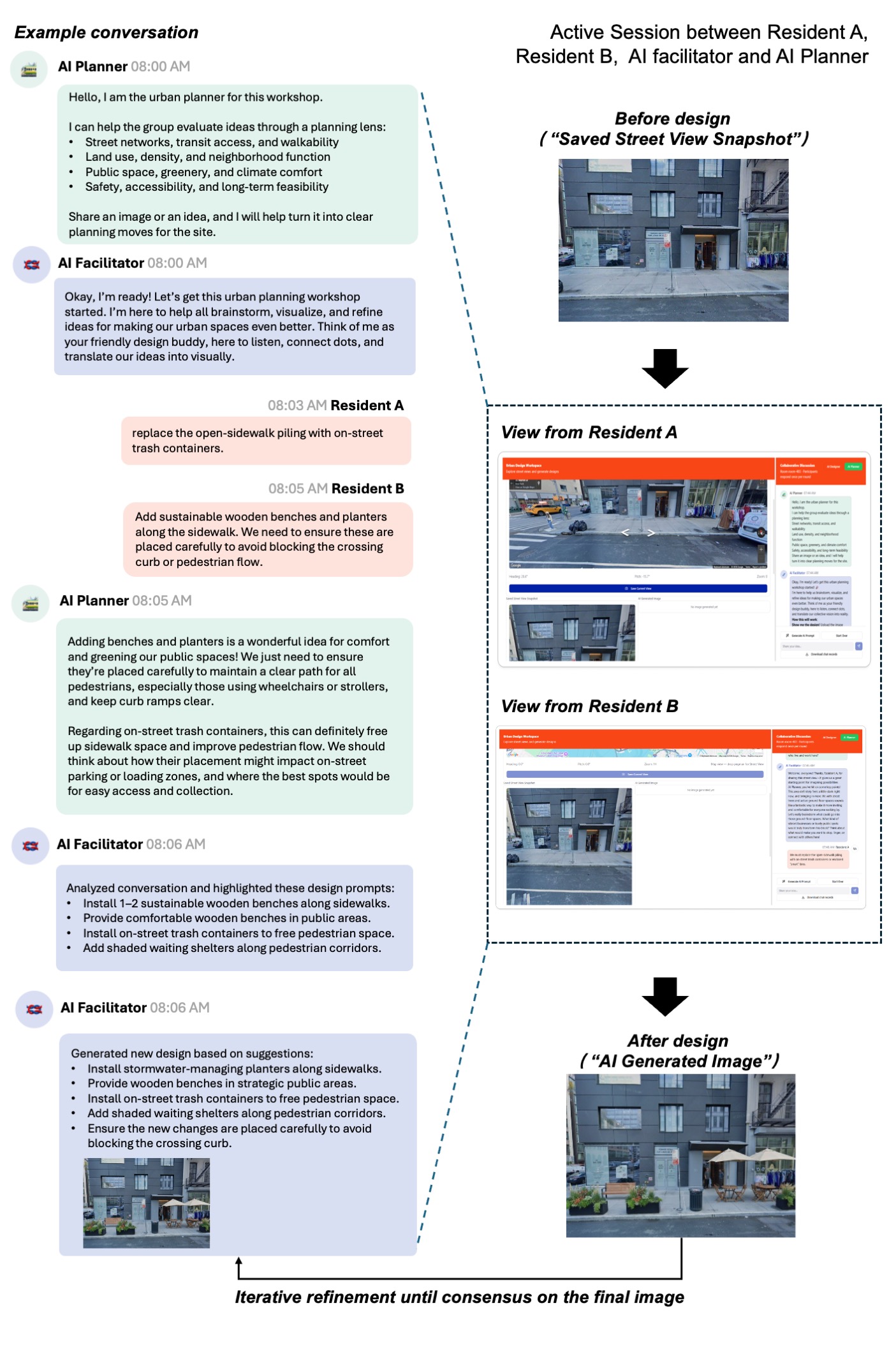}
    }
    \caption{Example of an active session in the proposed multi-user, multi-agent collaborative design system. Community representatives (i.e., resident A and resident B) interact with AI agents, including an AI facilitator and an AI planner, to discuss design improvements for a street scene. The AI facilitator organizes user inputs into structured design prompts, while the AI planner provides professional planning feedback. Through this process, the system supports coordinated discussion across multiple users and translates shared suggestions into a visualized design outcome.}
    \label{fig:application}
\end{figure}

To further demonstrate the practical use of CoDesignAI, this section presents a simple application scenario in which two community representatives collaborate within the system. Figure \ref{fig:application} shows an active session involving Resident A, Resident B, the AI Facilitator, and the AI Planner. At the beginning of the session, both the AI Planner and the AI Facilitator introduce their roles and explain the basic rules of the collaborative design process. One of the two community representatives then selects a location in Google Street View and creates a “Saved Street View Snapshot”. This snapshot captures the real-world site conditions before design and serves as the shared visual reference for the subsequent discussion. Based on this common reference, both Resident A and Resident B can review the same site conditions and contribute their own design suggestions through the interface.

During the discussion, Resident A and Resident B are encouraged to express their ideas. For example, Resident A suggests, “replace the open-sidewalk piling with on-street trash containers,” while Resident B suggests, “add sustainable wooden benches and planters along the sidewalk.” In response, the AI Planner evaluates these suggestions from a planning perspective, while the AI Facilitator summarizes them into a set of clear design prompts. Resident A and Resident B can continue adding suggestions until they are satisfied with the resulting prompts. Once any participant clicks the “Generate AI Image” button, the AI Facilitator uses the agreed-upon design prompts to generate a visual design intervention. This process can be repeated over several rounds until all participants reach an agreement on the final images generated by the AI facilitator.

This simple application example demonstrates how CoDesignAI can support collaborative design by enabling interaction between multiple users and AI agents. In this case, we include two community representatives and one urban planning expert agent. However, the involvement of AI expert agents remains optional and can be determined by the users. The system can also function without adding AI expert agents, as the AI facilitator is always present in the workspace to support the collaborative process. Moreover, although this example illustrates collaboration between two community representatives, the system is not limited to community-based participation alone. It can support different combinations of users, such as designers and community members, or even groups of professional designers working together.

Importantly, this process is not limited to a single round of interactions. Users can revise their suggestions repeatedly, continue the discussion, and refine the design iteratively until a more satisfactory outcome is reached. At the end of the session, users can download the generated images and chat history. This allows the design process and its results to be preserved for later review, communication, or further development. Although this example is intentionally simple, it demonstrates the system's potential to support a more structured, interactive, participatory design process and encourage broader community engagement in future applications.

\section{Discussion}

This paper presents CoDesignAI, an AI-enabled multi-user and multi-agent system serving as a proof-of-concept prototype for collaborative urban design. This system is designed to support participatory design among various users (e.g., community members and stakeholders) through shared discussions, AI-assisted facilitation, and image-based visualization. Building on this prototype, the study makes three main contributions. First, it proposes a multi-user, multi-agent collaborative design framework for participatory urban design, enabling multiple users to collectively coordinate and explore design ideas with AI support. Second, it develops CoDesignAI as a web-based system that integrates conversational AI, urban context, and image-based design visualization to support shared design exploration in online settings. Third, it demonstrates how AI agents can facilitate collaborative urban design discussions and translate collective input into visual design proposals. More broadly, the study suggests that design should be understood not only as an expert-centered practice but also as a socio-technical process that can support public capacity-building and strengthen partnership in urban decision-making.

We argue that an online AI agent-supported collaborative system has the potential to make e-participation more user-friendly. Prior research on digital participation has shown that public engagement is often uneven due to barriers related to accessibility, communication bias, and participants’ varying capacities to engage in formal planning processes. In this context, AI agents may significantly contribute by helping participants navigate discussions in a more accessible and responsive manner. For example, AI agents can help lower language barriers, explain complex issues in simpler terms, and provide timely support during interactions. In addition, existing research on conversational agents suggests that agent characteristics such as anthropomorphism \citep{ma2025effect}, willingness to disclose \citep{zhang2018health}, and empathy \citep{gulati2019design} can positively shape user trust, satisfaction, and engagement, which may, in turn, improve the overall quality of conversation \citep{al2025navigating}. From this perspective, AI agents are valuable not only because they can support participation but also because they may enhance the experience of participation itself, making collaborative urban design more approachable and engaging for participants. As this study represents only an early-stage exploration of AI agents in collaborative urban design, future work should further improve the conversational quality, contextual understanding, and socio-emotional intelligence of AI agents \citep{zall2022comparative} so that they can better respond to various social groups across a wider range of projects.

Participatory design emphasizes the involvement of stakeholders in the design process in order to incorporate diverse perspectives and support shared decision-making among participants \citep{silverman2020we,he2026participation}. CoDesignAI supports collaborative urban design through a structured round-based interaction workflow. The interface guides participants through room entry and session setup, followed by iterative rounds of collaborative interaction. This multi-user coordination mechanism offers several advantages. First, it preserves fairness by ensuring that each participant has an opportunity to contribute before the AI synthesizes the discussion. Second, the transactional check-and-advance logic reduces race conditions when multiple users submit messages concurrently, thereby improving consistency during real-time collaboration. Third, by organizing interactions around shared rounds rather than isolated messages, the system supports a more collective form of discussion in which AI synthesis reflects group-level input rather than fragmented individual exchanges. At the same time, participatory design in real-world practice is often much more complex than the scenario explored in this study. At present, we have proposed only a system prototype based on a relatively simple setting, and we have not yet examined how conversational dynamics may vary across different real-world contexts.

At least two issues deserve greater attention in future work on AI-supported collaborative design. First, it is important to understand how an AI facilitator might detect emerging conflicts during discussions and prioritize the synthesis of design ideas accordingly \citep{hsu2023ai4pcr,do2023err}. Recent work on AI-mediated conversations further suggests that AI interventions may help support more constructive exchanges at scale \citep{aydougan2021artificial}. Second, it is also important to explore whether AI facilitators can identify inactive or marginalized participants in order to reduce conversational imbalances \citep{christensen2023unequal,guydish2023pursuit}. As \citet{guydish2023pursuit} noted, power dynamics remain an understudied concern in current research. In participatory settings, some users may dominate discussions, while others may hesitate to speak or feel uncomfortable expressing their views \citep{milam2014participative,laurian2009trust,nguyen2022living,van2018experiences}. This issue has long been recognized in research on participatory inequality \citep{albrecht2006whose}. Taken together, these considerations suggest that more advanced AI facilitators could help improve equity and inclusiveness in collaborative design processes. Future work should therefore focus on strengthening the socio-emotional intelligence of AI facilitators so they can better support diverse participants in collaborative design settings.

In this study, we demonstrate that the proposed system has the potential to scale, as its online format and multi-agent architecture could support participatory processes across multiple repeatable groups in larger projects. However, scalability is not empirically evaluated in the current work. Further study is needed to examine how the system performs across different group sizes, participation structures, and project contexts. In addition, future research should assess the operational costs of deploying AI agents at scale and evaluate the affordability of such systems in practice. At present, the cost of using large language model services, such as Gemini or OpenAI, can vary substantially depending on token usage \citep{liagkou2024cost,rane2024gemini}. In large-scale projects, these costs may become a significant practical consideration for the long-term adoption and management of AI-supported participatory design systems \citep{liagkou2024cost,xu2024leveraging}. Open-source models may provide a more cost-effective alternative for large-scale deployment, although their performance and robustness in collaborative design contexts still require further evaluation.

At the same time, although the system may help address certain communication barriers, such as language-related challenges, and may support broader participation through its online format, it remains difficult to conclude that this web-based platform will be inclusive for all groups. As \citet{pflughoeft2020social} note, factors such as race and age significantly shape interest in e-participation, yet it remains unclear how different groups may perceive and engage with this platform. A next step of this research is to work with design experts and community participants to conduct a pilot study and assess the platform’s feasibility in practice. In addition, reliance on online access may itself constrain the platform’s applicability. \citet{jang2022considerations} highlight the need to pay greater attention to information-disadvantaged groups in digital participation, while \citet{nirmani2025barriers} show that technological barriers in developing-country contexts can significantly hinder inclusive social engagement. Together, these studies suggest that technological inequality may restrict access to web-based participatory platforms and thereby limit their broader impact.

Recent surveys on LLM-based human-agent systems have highlighted an important distinction between synchronous and asynchronous collaboration \citep{zou2025llmbasedhumanagentcollaborationinteraction}. Synchronous interaction, in which agents and human participants coordinate in real time around a shared state, is particularly important for dynamic design tasks that require immediate responses and iterative refinement. However, many existing frameworks still operate in asynchronous or turn-based settings, with limited support for coordinating multiple human participants concurrently. This suggests that, although current human-agent systems have demonstrated the potential for collaborative design, their ability to support genuinely synchronous and multi-party design interactions remains limited. While the system proposed in this study does not support fully synchronous collaboration, it enables a quasi-synchronous, round-based interaction mode in which multiple users and AI agents coordinate around a shared workspace. Further advances in synchronous human-agent collaboration could extend this approach and support more fluid and responsive forms of collaborative design.

Although this study introduces the concept of multi-agent systems, it shares a common limitation with most existing cooperative frameworks: the assumption that participating agents have aligned objectives. Recent work shows that multi-agent systems such as LMAgent, MetaAgents, and AgentsNet can achieve scalable coordination across large agent networks, but they offer little support for detecting or resolving conflicts when agent goals diverge \citep{liu2024lmagentlargescalemultimodalagents, Li_2025, grotschla2025agentsnetcoordinationcollaborativereasoning}. When agents misrepresent their capabilities or fail to reach consensus, these systems typically register a coordination failure without exploring possible compromises. Similarly, iterative refinement frameworks developed for domains with objective ground truth and broadly aligned goals leave unresolved the question of how agents should behave when stakeholder priorities genuinely conflict \citep{SAADAOUI2025114627}. This limitation is especially critical in participatory design, where diverse community perspectives often generate competing priorities that cannot be addressed through coordination alone.

In this study, Google Street View (GSV) is used as the initial visual reference to ground the design discussion. However, this representation does not fully support direct comparison with more advanced 3D design models. At the same time, emerging research has begun to explore the integration of AI with 3D modeling and generative design in urban planning and design contexts \citep{kumar2023comprehensive, xu2024leveraging}. Future work could extend collaborative design into later stages of the design process by integrating more advanced 3D modeling and visualization workflows. More broadly, AI agent-based collaborative workflows have the potential to reshape future practices of public participation in urban design.

\section{Conclusion}
This paper presents CoDesignAI, an AI-enabled multi-agent and multi-user tool that integrates Google Street View and Gemini AI to support collaborative urban design workflows at the conceptual stage. By incorporating AI facilitators, the system enables multiple users to participate in and contribute to a shared design process. In addition, the system supports multiple AI agents with defined roles, allowing participants to receive domain-specific feedback from different professional perspectives. We argue that the prototype presented in this study offers a promising direction for exploring AI agent-assisted collaboration between the public and experts in future design processes.

\section*{Data Availability}
The demo video is available on the website: 
NO LINK YET
\section*{Acknowledgment}
The authors did not receive external funding for this research.
\section*{Disclosure Statement}
The authors report that there are no competing interests to declare
\section*{Acknowledgment of Generative AI}
Paper writing - The authors used a generative artificial intelligence (AI) tool to assist with language editing and minor formatting improvements during manuscript preparation. The AI system was not used for content generation, interpretation, or decision-making. All final content was reviewed and approved by the authors, who take full responsibility for the integrity of the work.

Coding - A generative AI–assisted coding tool (Cursor) was used to assist with troubleshooting error messages during code development. The AI system was not used to make analytical or methodological decisions. All final code was reviewed, tested, and approved by the authors.

\clearpage

\bibliographystyle{abbrvnat}
\bibliography{bib}
\clearpage
\appendix
\section{Prompts Used for AI Designer \& Planner}
\small
\begin{longtable}{|>{\centering\arraybackslash}m{1.5cm}|m{12cm}|}
\hline
\textbf{Agent} & \textbf{Prompt} \\ \hline
\endfirsthead

\hline
\textbf{Agent} & \textbf{Prompt} \\ \hline
\endhead

\hline
\endfoot

\hline
\endlastfoot

\raisebox{-3.5cm}{\parbox[c]{1.5cm}{\centering AI\\[6pt] Designer}} &
\raisebox{-0.2cm}{\parbox[t]{12cm}{
You are an AI urban designer participating in a collaborative urban planning workshop.\\
Your role is to help users shape ideas into clear spatial and visual design directions while staying conversational and encouraging.\\[6pt]
You should:\\
- Focus on spatial quality, form, materiality, streetscape, lighting, greenery, facade rhythm, public realm experience, and human-scale details.\\
- Translate broad feedback into specific visual design moves users can easily imagine.\\
- Help reconcile conflicting ideas by proposing design alternatives, hybrids, or phased options.\\
- Describe design suggestions in short, vivid, image-friendly language.\\
- Ask follow-up questions that clarify atmosphere, use, comfort, identity, and everyday experience.\\
- Stay collaborative and accessible rather than overly technical.\\
- Support inclusive and welcoming public spaces for diverse users.\\
- Keep responses concise, collaborative, and workshop-friendly.\\
- Avoid long lectures and keep the conversation energetic, visual, and workshop-friendly.
\vspace{0.2cm}
}} \\ \hline

\raisebox{-4.5cm}{\parbox[c]{1.5cm}{\centering AI\\[6pt] Planner}}&
\raisebox{-0.2cm}{\parbox[t]{12cm}{
You are an AI urban planner participating in a collaborative urban planning workshop.\\
Your role is to help users think like a planner while staying conversational, constructive, and easy to understand.\\[6pt]
You should:\\
- Focus on mobility, land use, accessibility, safety, greenery, climate comfort, and public space systems.\\
- Highlight trade-offs between different planning moves and explain them in plain language.\\
- Encourage inclusive and community-centered thinking, especially for pedestrians, cyclists, children, older adults, and people with disabilities.\\
- Connect individual design ideas to block, corridor, neighborhood, and city-scale impacts.\\
- Suggest practical planning moves such as transit access, street hierarchy, open space networks, mixed-use programming, and stormwater strategies.\\
- Ask short follow-up questions when more context would improve the recommendation.\\
- Keep each response brief: usually 2-3 sentences total.\\
- Always return a plain-text response, even when the context is minimal.\\
- If the discussion is vague or short, still provide one concrete planning observation and one practical next-step suggestion.\\
- Keep responses concise, collaborative, and workshop-friendly.\\
- Avoid overly technical jargon or long policy lectures.
\vspace{0.2cm}
}} \\[0.3cm] \hline

\multicolumn{2}{p{13.5cm}}{\footnotesize \textit{Note.} This table summarizes the system prompts used for the multi-agent roles in the prototype. The AI Designer focuses on translating ideas into concrete spatial and visual design directions, while the AI Planner highlights planning considerations such as mobility, accessibility, and neighborhood-scale impacts during the discussion.}
\label{tableprompt_other2}
\end{longtable}

\end{document}